\documentclass[twocolumn,showpacs,superscriptaddress,pra]{revtex4}
\usepackage{amsbsy}
\usepackage{amsfonts}
\usepackage{amssymb}
\usepackage{amsmath}
\usepackage{graphicx,color}
\usepackage{graphicx}

\newcommand{\ie}{{\it{i.e.}}}
\newcommand{\etal}{{\it{et al.}}}
\newcommand{\refeq}[1]{Eq.~(\ref{#1})}
\newcommand{\Tr}{\mathrm{ Tr }}
\newcommand{\oz}{\Omega_0}

\newcommand{\cz}{C_0}
\newcommand{\chb}{{\cal H}_B}

\newcommand{\bqh}{{\cal B}{\cal (H)}}

\newcommand{\oket}[1]{\left| #1 \right)}
\newcommand{\obra}[1]{\left( #1 \right|}
\newtheorem{prop}{Proposition}

\begin{document}
\title{Almost all quantum states have non-classical correlations}

\author{A. Ferraro}
\affiliation{ICFO-Institut de Ciencies Fotoniques, Mediterranean
Technology Park, 08860 Castelldefels (Barcelona), Spain}
\author{L. Aolita}
\affiliation{ICFO-Institut de Ciencies Fotoniques, Mediterranean
Technology Park, 08860 Castelldefels (Barcelona), Spain}
\author{D. Cavalcanti}
\affiliation{ICFO-Institut de Ciencies Fotoniques, Mediterranean
Technology Park, 08860 Castelldefels (Barcelona), Spain}
\affiliation{Center for Quantum Technologies, University of Singapore, Singapore}
\author{F. M. Cucchietti}
\affiliation{ICFO-Institut de Ciencies Fotoniques, Mediterranean
Technology Park, 08860 Castelldefels (Barcelona), Spain}
\author{A. Ac\'in}
\affiliation{ICFO-Institut de Ciencies Fotoniques, Mediterranean
Technology Park, 08860 Castelldefels (Barcelona), Spain}
\affiliation{ICREA-Instituci\'o Catalana de Recerca i Estudis
Avan\c cats, Lluis Companys 23, 08010 Barcelona, Spain}

\begin{abstract}
Quantum discord quantifies non-classical correlations in a quantum system including those not captured by entanglement. Thus, only states with zero discord exhibit strictly classical correlations. 
We prove that these states are negligible in the whole Hilbert space: typically a state picked out at random has positive discord; and, given a state with zero discord, a generic arbitrarily small perturbation drives it to a positive-discord state. 
These results hold for any Hilbert-space dimension, and have direct implications on quantum computation and on the foundations of the theory of open systems. In addition, we provide a simple necessary criterion for zero quantum discord. Finally, we show that, for almost all positive-discord states, an arbitrary Markovian evolution cannot lead to a sudden, permanent vanishing of discord. 
\end{abstract}

\date{\today}


\maketitle
The emergence of quantum information science motivated a major
effort towards the characterization of entangled states, generally believed to be an essential resource for quantum information tasks that outperform their classical
counterparts. In particular, 
the geometry of the sets of entangled/non-entangled states received much
attention \cite{BZ} -- starting from the fundamental result
that the set of separable (non-entangled)
states has non-zero volume in a finite dimensional Hilbert space \cite{ZHSL}. In other words, separable states are not at all negligible,
which has direct implications  on some
implementations of quantum computing \cite{B+} and on the definition of
entanglement quantifiers \cite{VT}.

\par Apart from entanglement, quantum states display other correlations \cite{OZ,HV,Terhal&DiVincenzo} not present in classical systems 
(meaning, here, systems where all observables commute). Aiming at capturing such correlations, Ollivier and Zurek introduced the quantum
discord \cite{OZ}.
They showed that only in the absence of discord there exists a
measurement protocol that enables distant observers to extract
all the information about a bipartite system without perturbing it. This
completeness of local measurements is featured by any classical
state, but not by quantum states, even 
some separable ones. Thus,  zero discord is a necessary condition for  only-classical  correlations. 

\par Very recently, quantum discord has received increasing attention \cite{Luo&Wu,DSC,DThesis,SL,PHH,W+,L+,others}.
A prevailing observation in all results obtained so far is that the absence or presence of discord 
is directly associated to non-trivial properties of states.
Thus,  it is natural to question {\it how typical are positive-discord states?}.
Here we prove that a particular subset of states that contains {\it the set of zero-discord states, has measure zero and is nowhere dense}. 
That is, it is topologically negligible:
typically, every state picked out at random has positive discord; and given a state with zero discord, a generic (arbitrarily small)
perturbation will take it to a state of strictly positive discord.  
Remarkably, these results hold true for any Hilbert space dimension and are thus in contrast with expectations based on the structure of entangled states \cite{ZHSL}: while the set of separable states has positive volume, the set of only-classically correlated states does not.
In addition, we provide a novel necessary condition, of very simple evaluation, for zero quantum discord. 
With this tool we suggest a schematic geometrical representation 
of the set of zero discord, and study the open-system dynamics of discord. 
We find that for almost all states of positive discord, the interaction with any (non-necessarily local) 
Markovian bath can {\em never}  lead either to a sudden, permanent  vanishing of discord, nor to one lasting a finite time-interval. 
In strong contrast to entanglement -- which typically vanishes 
suddenly and permanently at a finite time \cite{YE} --,
discord can only permanently vanish in the asymptotic inifnite-time limit, i.e.  at the steady state.

Our results have wide-range implications. First, from a fundamental perspective, they imply that
only-classically correlated states are  extremely rare in the space of all quantum states. 
Second, it has been recently discovered that an arbitrary unitary evolution for any system and  bath is described
(upon tracing the bath out) as a completely-positive map on the system if, and only if, system and bath are initially in a 
zero-discord state \cite{SL}. In view of the rarity of zero-discord states, the fundamental recipe 
``unitary evolution $+$ partial trace" is now in conflict with complete positivity -- one of the most basic and fundamental
requirements that physical evolution is demanded to fulfill \cite{OpenSystems} -- for almost all quantum states.
Another interesting fact is that quantum discord is present in typical instances of a mixed-state quantum computation 
\cite{KL}, even when entanglement is absent \cite{DSC,DThesis, L+}. This led Datta et al. \cite{DSC,DThesis}
to suggest that discord might be the resource responsible for the quantum speedup in this computational model.
If the mere presence of discord was by itself responsible of some speedup, then our results would imply that 
almost all quantum states are useful resources. Furthermore, Piani et al.  \cite{PHH} introduced a new task -- local broadcasting -- to operationally distinguish among different varieties of  states with zero quantum discord. They showed that only some zero-discord states can be locally broadcasted,  which  
-- according to us -- now means hardly any quantum state.
Also, our general results on the Markovian dynamics of discord
complement and generalize the specific results reported in Refs.~\onlinecite{W+}. 
There, for particular cases of local channels and two-qubit systems, 
discord was never observed to vanish permanently at a finite time. 
As said, we prove the generality of this behavior. 
Finally, our results also apply to quantifiers of quantum correlations other than discord.

{\em Quantum Discord}.--
Consider a bipartite system in a composite Hilbert space ${\cal H}={\cal H}_A\otimes{\cal H}_B$, 
of dimension  $d=d_A\times d_B$, with $d_A=\mathrm{dim}({\cal H}_A)$ and $d_B=\mathrm{dim}({\cal H}_B)$, 
respectively. Given a quantum state $\rho\in\bqh$ (where $\bqh$ denotes the set of bounded, positive-semidefinite operators 
o n $\cal H$  with unit trace), the von Neumann mutual information $I_{AB}$ 
between $A$ and $B$ is defined as
\begin{equation}
I_{AB}(\rho)\doteq S(\rho_A)+S(\rho_B)-S(\rho),
\label{minfo1}
\end{equation}
where $S(\rho)=-\Tr[\rho \log \rho]$ is the von Neumann entropy and $\rho_{A,B}=\Tr_{B,A}[\rho]$. Mutual information \eqref{minfo1} quantifies the total amount of correlations in quantum states \cite{HV}.

A classically equivalent definition of mutual information is $S(\rho_B)-S(\rho_{B|A})$, where $\rho_{B|A}$ is the state of
$B$ given a measurement in $A$. Thus, classical mutual information quantifies  the  decrease in ignorance (gain of information) 
about subsystem $B$ upon  local measurement on $A$.  Let us now consider a measurement consisting of  (non-necessarily orthogonal)
one-dimensional measurement elements $\{{M_j}\}$ on ${\cal H}_A$. We can write the state of system $B$ conditioned on the 
outcome $j$ for $A$ as
$
\rho_{B|j}=\Tr_A[M_j \rho] /p'_j ,
$
where the probability of outcome $j$ is given by $p'_j=\Tr [\rho M_j]$. 
By optimizing over the measurement set $\{{M_j}\}$, one can define 
\begin{equation}
J_{AB}(\rho)\doteq S(\rho_B)-\mathrm {min}_{\{M_j\}} \sum_j p'_j S(\rho_{B|j}),
\label{minfo2}
\end{equation}
which quantifies the classical correlations in $\rho$ \cite{HV}. 

\par Despite both definitions for the mutual information being equivalent for classical systems, 
the quantum generalizations $I_{AB}$ and $J_{AB}$ in general do not coincide: Their discrepancy defines the discord:
\begin{equation}
D_{AB}(\rho)\doteq I_{AB}(\rho)-J_{AB}(\rho).
\label{disco}
\end{equation}
Notice that quantum discord is always non-negative and it is asymmetric with respect to $A$ and $B$ \cite{OZ}. 

{\em Null-discord states}.--
Let us denote by $\oz$ the set composed of all states with zero discord:
\begin{equation}
\oz\doteq \{ \rho\in\bqh \;\; \mathrm{s.t.} \;\; D_{AB}(\rho)=0\}.
\label{oz}
\end{equation}
The members of this set are characterized \cite{OZ,DThesis} by being invariant under von
Neumann measurements on $A$ in some orthonormal basis $\{{\Pi}_j\}$, that is
\begin{equation}
\rho\in\oz \Longleftrightarrow\exists\;\; \{{{\Pi}_j}\} \;\; \mathrm{s.t.} \;\;  \rho=\sum_{j=1}^{d_A} {{\Pi}_j}\rho{{\Pi}_j}.
\label{oziff}
\end{equation}
This implies that the set $\{{{\Pi}_j}\}$ defines a basis of ${\cal H}_A$ 
with respect to which $\rho$ is block diagonal \cite{DThesis}:
\begin{equation}
\rho\in\oz \Longleftrightarrow\exists\;\; \{{{\Pi}_j}\} \;\; \mathrm{s.t.} \;\;\rho=\sum_{j=1}^{d_A} p_j {{\Pi}_j}  \otimes \sigma_j \; ,
\label{bdiag}
\end{equation}
where $\sigma_j$ are quantum states in ${\cal  B}({\cal  H}_B)$ and $\{p_j\}$ defines a probability distribution.

\par The characterization of $\oz$ presented just above is not practical in the sense that one has to check for the existence of a measurement basis for which conditions \eqref{oziff} and \eqref{bdiag} are satisfied. With this motivation, we derive a sufficient condition for positive quantum discord  that is {\it basis-independent}. From condition \eqref{bdiag}, and denoting by $[,]$ the commutator, it follows that
\begin{prop}\label{propo1}
If $\rho\in\oz$ then 
\begin{equation}
[\rho,\rho_A\otimes \openone_B]=0,
\label{oz_nec}
\end{equation}
where $\openone_B$ is the identity operator on $\chb$. Hence, $[\rho,\rho_A\otimes \openone_B]\neq0$ implies that $D_{AB}(\rho)>0$.
\end{prop}
The converse, however, is not true: there are some  states with positive discord that commute with their reduced ones.
States of  interest like all pure maximally-entangled states are an example. 
Let us introduce the auxiliary set $\cz$ of all states satisfying \refeq{oz_nec}:
\begin{equation}
\cz\doteq \{ \rho\in\bqh \;\; \mathrm{s.t.} \;\; [\rho,\rho_A\otimes \openone_B]=0\}.
\label{cz}
\end{equation}
One has that $\oz\subset\cz$. We prove next that $\cz$ has measure zero and is nowhere dense, thereby implying the same properties
for $\oz$ \cite{note_other}. 

{\em $\cz$ has measure zero}.-- The key observation here is that Eq. \eqref{oz_nec} imposes a non-trivial constraint on $\rho$ that
confines it to a lower dimensional subspace of $\bqh$. This already suggests that the volume of $\cz$ in $\bqh$ is zero, a proof of which we sketch next (a detailed proof is given in Appendix  \ref{measure_zero}). Consider a generic state $\rho\in\bqh$ expressed for example in an orthogonal basis given by the tensor product
between the traceless generators of the group $SU(d_A)$ and those of $SU(d_B)$. In this basis, 
the calculation of commutator (\ref{oz_nec}) is straightforward and gives a set of implicit constraints on a state to belong to 
$\cz$. These constraints can be inverted to obtain an explicit differentiable parametrization of the set $\cz$ which uses strictly
fewer independent real parameters than the ones needed to parametrize $\bqh$. 
Since a differentiable parametrization of a set measure zero is also measure zero, $\cz$ has measure zero in $\bqh$.

{\em $\cz$ is nowhere dense}.--
The set $\cz$, apart from being of zero measure,
is also nowhere dense, two concepts a priori independent. 
A set $\cal A$ is called
nowhere dense (in $\cal X$) if there is no neighborhood in $\cal X$ on
which $\cal A$ is dense. Equivalently, $\cal A$ is said to be nowhere
dense if its closure has an empty interior. In particular, this
implies that within an arbitrarily-small vicinity of any state  that belongs to $\cz$ ($\oz$) there are always states  
out of $\cz$ ($\oz$). 
Let us next observe that $\cz$ is closed. This follows from the fact that the 
function $f(\rho)=[\rho,\rho_A\otimes \openone_B]$ is a continuous map and the zeros of a continuous map form a closed set.
Since any closed set of measure zero is nowhere dense, this suffices to conclude that $\cz$  is nowhere dense (see  Appendix \ref{closed_set}). Being both closed and nowhere dense implies in particular that a generic perturbation
of a state inside the set will drive it not just to a state outside, but to an entire region (an open set) outside of it. 

{\em Geometry of the set of zero quantum discord}.--
First, let us observe that $\oz$ is not a convex set. In fact, an arbitrary convex mixture  between two states $\rho_1$
and $\rho_2$  that are block diagonal in incompatible local bases is typically not block diagonal. 
On the other
hand, if one mixes  states  block diagonal in the same local basis, then the resulting state is necessarily block diagonal (in the 
same basis), and therefore belongs
to $\oz$.  In
particular,  every state of zero discord is connected to the maximally mixed state $\openone/d$, as the latter is trivially block diagonal in any local basis. 
This already shows us that the set $\oz$ is connected. 
From a  geometrical viewpoint, this means that when moving rectilinearly from every state in $\oz$ towards $\openone/d$, only states in $\oz$ are encountered. Accordingly, the segment from every state out of $\oz$ to $\openone/d$ is exclusively composed of states out of $\oz$
(being the two-qubit Werner state an instructive and simple example \cite{OZ}).
All in all, this leaves us with some sort of star-like hyper-structure for $\oz$ (with 
$\openone/d$ at the center), represented in Fig. 1.

Some details of the set have been sacrificed in the figure for the sake of clarity.  For example, the tips of the star rays are pure separable states, always at the border of $\bqh$, even though some of them are shown in its interior. Also, all the rays are connected not only through  $\openone/d$, but also by (non-convex) continuous trajectories induced by  local unitaries. This nevertheless, as we already know, lies fully in a lower-dimensional subspace without volume and  is not represented  in Fig. 1. The picture should thus not be taken as rigorous but just as a pictorial representation to illustrate the main features  of $\oz$.

The geometrical notion of moving rectilinearly toward $\openone/d$ corresponds to the dynamical process of global depolarization (global white noise). 
From the above considerations,  it is clear now that global depolarization can never induce finite-time vanishing of positive  discord. It only induces the disappearance of discord in the asymptotic infinite-time limit, when $\openone/d$ is actually reached.
In fact, given the singular geometry of $\oz$ suggested here, it seems highly unlikely that a noisy dynamical evolution inducing a smooth trajectory
in $\bqh$ is able to take a state outside of  $\oz$ into its interior, and to keep it there permanently. 
This is what we discuss next.  
\par \begin{figure}[htbp] 
  \centering
  \includegraphics[width=3.1in]{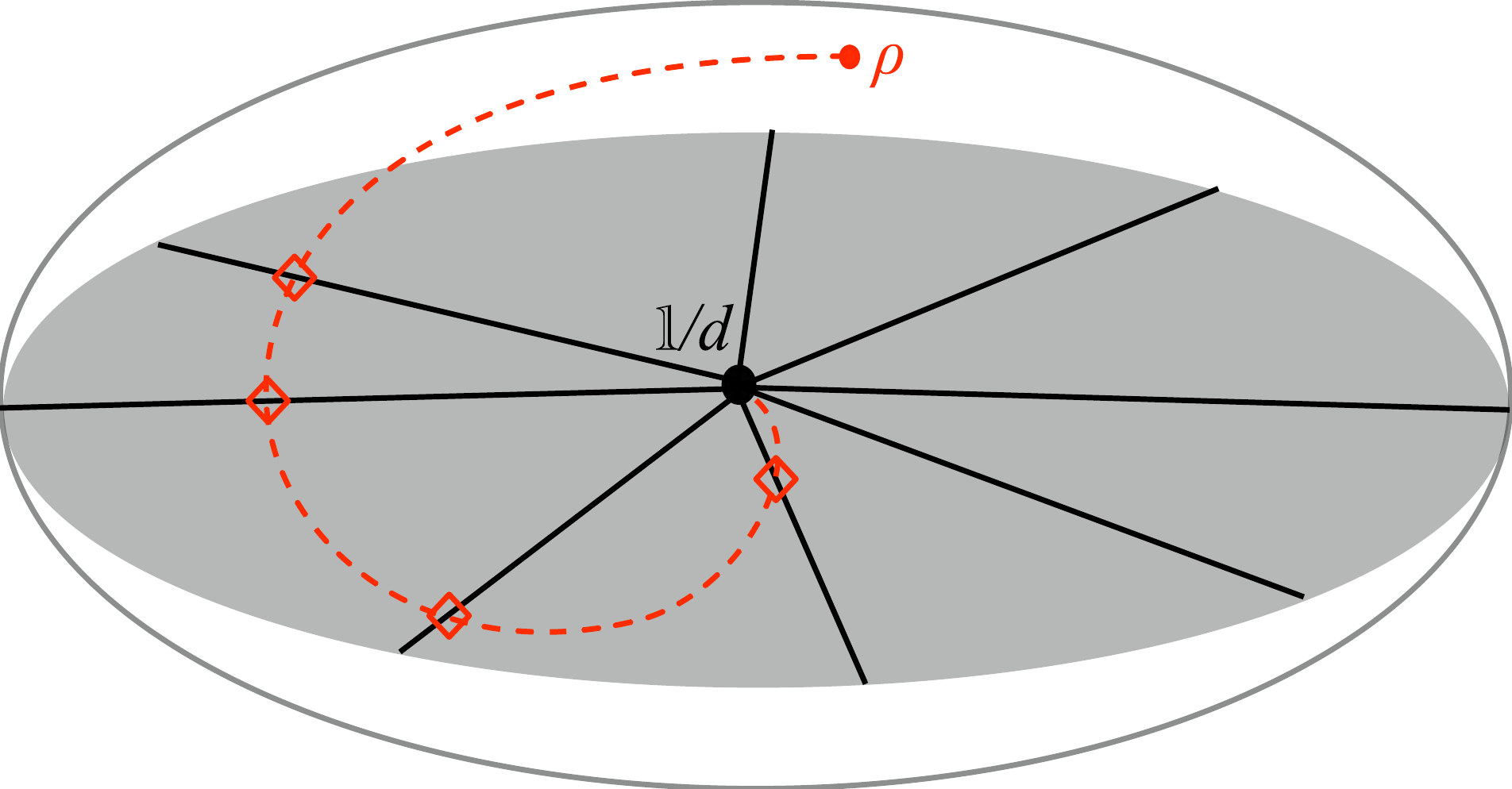} 
  \caption{(Color online). Schematic 2D representation of the set $\oz$ of states with zero discord (dark lines).
  The set of all possible states $\bqh$ (enclosing ellipse) contains the set of separable ones, depicted in grey, with the maximally mixed state $\openone/d$ in its
  center. All block diagonal states, including pure separable states at the border of $\bqh$, compose $\oz$, and can be connected to $\openone/d$ through
  states in $\oz$. Arbitrary states in $\oz$, however, cannot in general be combined to form a state in $\oz$. The whole of $\oz$ lives in a lower dimensional subspace of $\bqh$. The dynamical trajectory of an arbitrary state $\rho$ caused by a Markovian bath is represented in dashed-red. In this example the trajectory leads towards  $\openone/d$. During its evolution, the evolved state can only cross $\oz$ a finite number of times, and permanent vanishing of discord cannot happen before the  infinite-time limit, at the stationary state. }
  \label{fig}
\end{figure}

{\em Open-system dynamics of discord}.-- 
We now show for any state $\rho\notin\cz$ -- that is, for almost all (positive-discord) states -- that the interaction with any 
(non-necessarily local) Markovian bath can never lead to a sudden permanent vanishing of discord. Unless the asymptotic infinite-time limit 
is reached, a Markovian map can take $\rho$ through the singular set $\cz$ (and therefore also through $\oz$) at most a finite 
number of times, equal to $\tilde{d}_{\lambda}(\tilde{d}_{\lambda}-1)/2-1$, where $\tilde{d}_{\lambda}$ is the number of different eigenvalues of the map. 

\par Consider the system interacts with a generic (non necessarily local) bath during an arbitrary time $\tau$.
We describe the evolution of the system with a  completely-positive, 
trace-preserving map $\Lambda_\tau:\bqh\longrightarrow\bqh$. In what follows we use the notation
of Ref. \cite{Caves}. The map
$\Lambda_\tau$ can be written in its (diagonal) spectral decomposition, 
$\Lambda_\tau=\sum\lambda_{i} \oket{\mu_{i}} \obra{\nu_{i}}$, where $\lambda_{i}$,  $\oket{\nu_{i}}$ and $\oket{\mu_{i}}$ are  respectively the
eigenvalues, left and right eigenoperators of the map,  $\Lambda_\tau\oket{\mu_{i}}\equiv\lambda_{i}\oket{\mu_{i}}$ and $\obra{\nu_{i}}\Lambda_\tau\equiv\lambda_{i}\obra{\nu_{i}}$. For a general map, $\oket{\nu_{i}}$ and $\oket{\mu_{i}}$ span two non-orthogonal complete bases of $\bqh$ and satisfy the conditions $\obra{\nu_{i}}\mu_{j})\equiv\delta_{ij}$ and $\obra{\nu_{i}}\nu_{j})\neq\delta_{ij}\neq\obra{\mu_{i}}\mu_{j})$, where $\delta_{ij}$ is the Kronecker delta and where $(X|Y)$ is nothing but the Hilbert-Schmidt inner product: $(X|Y)\equiv\Tr[X^{\dagger}.Y]$. In addition, these maps are always contractive, that is, $|\lambda_{i}|\leq1\forall\ i$ and $|\lambda_{i}|=1$ for at least one $i$.  For the specific case of normal maps (those commuting with their adjoints) the left and right eigenoperators coincide and the basis they span becomes orthonormal. Also, since we are interested in maps that describe some decoherence process, we assume that  $|\lambda_{i}|<1$ for at least one $i$, for the case  $|\lambda_{i}|=1\forall\ i$ corresponds to the  case of unitary evolution of the composite system.

\par We consider now all maps $\Lambda_t$ that can be expressed as the successive composition of $n$ times $\Lambda_\tau$: $\Lambda_t=\sum\lambda_i^n\oket{\mu_i} \obra{\nu_i}$, with $t=n\tau$. All Markovian maps fall into this category. From a strictly mathematical viewpoint, it is possible that some of the eigenvalues of $\Lambda_\tau$ are null. Nevertheless, since the initial condition $\Lambda_{\tau=0}\equiv\openone$ must be satisfied, because of continuity there is always a sufficiently  small $\tau$ for which all eigenvalues are non-null. With this physically motivated observation 
in mind, we restrict our discussion to all maps such that $\lambda_i\neq0\ \forall\ i$. The initial state $\rho$ is expanded in the basis $\left\{\oket{\mu_i}\right\}$ as $\rho=\sum\rho_i \oket{\mu_i}$, with $\rho_i\equiv\obra{\nu_i}\rho)$, and after time $t$ it evolves to $\rho_t\equiv\Lambda_t(\rho)=\sum\rho_i \lambda_i^n
\oket{\mu_i}$. 
Now we can show that for a generic (positive-discord) initial state  $\rho$ such that $[\rho,\rho_A\otimes \openone_B]\neq0$,
there exists no  $t_s\in[0,\infty)$ such that $\rho_t\in\cz$  (and in particular such that $\rho_t\in\oz$) for all $t>t_s$.
We do it by {\it reductio ad absurdum}. Assume then the opposite is true.  This
means that there exists a state $\rho_t$ that satisfies $[\rho_t,{\rho_A}_t\otimes\openone_B]=0$, with ${\rho_A}_t\equiv\Tr_B[\rho_t]$, for all $t>t_s$. This, however, defines an {\em infinite} set of linearly independent equations (as many as $n>n_s\equiv t_s/\tau$), 
which can never be satisfied.  An analogous contradiction is obtained also if it is assumed  that $\rho_t$ satisfies $[\rho_t,{\rho_A}_t\otimes\openone_B]=0$ only during the finite-time interval $(t_s,t_s+\Delta t]$, with any $\Delta t>0$.
Furthermore, we prove that $\rho_t$ can enter $\cz$  (and in consequence also $\oz$) a maximum of 
$\tilde{d}_{\lambda}(\tilde{d}_{\lambda}-1)/2 - 1$ times, where $\tilde{d}_{\lambda}$ is the number of different eigenvalues  $\lambda_i$
(see Appendix \ref{Noaccording}).

{\em Discussion}.--
We have shown here that a random quantum state have strictly positive discord and a generic small perturbation of a state with zero discord will generate discord.
These results imply that only-classically correlated quantum states are extremely rare.
An interesting analogy can now be
established: almost all states possess discord just as
almost all {\it pure} states possess entanglement. This means
that the mere presence of positive quantum discord lacks {\it per se} informative
content (for example as a computational resource), being it a common feature of almost all quantum
states. Of course, this by no means
excludes the possibility that a more quantitative characterization of the discord gives valuable assessment of a
state's usefulness for some task.
In a future perspective, our results call for a better understanding of the conflict between the standard approach to open quantum systems and complete positivity of maps.  

A final comment about experimental implications. We have shown that states with zero discord are
(densely) surrounded by states with positive discord. As a consequence,
ruling out the presence of quantum discord is, strictly speaking,
experimentally impossible (unless further assumptions are taken).
The reason is that any measurement with a non-null error range is
compatible with a positive amount of discord. This is
in striking contrast to what happens for entanglement, whose presence can
instead be  strictly ruled out in experiments.  


We thank  A. Datta, R. Drumond, I. Garc\'\i a-Mata, M. T. Cunha, J. Wehr, A. Winter and W. Zurek for discussions; and the European QAP, COMPAS and PERCENT projects, the Spanish MEC FIS2007-60182 and Consolider-Ingenio QOIT projects,  the Generalitat de Catalunya, and Caixa Manresa for financial support.

\appendix

\section{The set $\cz$ has measure zero}
\label{measure_zero}

We express $\rho\in\bqh$  in the basis given by the traceless, orthogonal generators $\gamma_i^A\otimes\gamma_j^B$ of the product group $SU(d_A)\otimes SU(d_B)$ (we use the same notation as Eq. (5.2) of Ref. [K.~\.{Z}yczkowski and I. Bengtsson, arXiv: quant-ph/0606228]):
\begin{align}
\rho  =\frac{1}{d_Ad_B} & \left[\openone_{AB}
+\sum_{i=1}^{d_A-1}\tau_i^A\gamma_i^A\otimes\openone_B
+\sum_{j=1}^{d_B-1}\tau_i^B\openone_A\otimes\gamma_i^B
\right.
\nonumber \\
& \left.+\sum_{h=1}^{d_A-1}\sum_{k=1}^{d_B-1}\beta_{hk}\gamma_h^A\otimes\gamma_k^B
\right] \;.
\end{align}
The expression above maps the Hilbert space $\cal H$ to ${\mathbb R}^{d_0}$ ($d_0=d_A^2d_B^2-1$) via the parameters ${\tau_i^A}$, ${\tau_i^B}$, and $\beta_{hk}$. The partial trace of $\rho$ over $\chb$ gives:
\begin{equation}
\rho_A  =\frac{1}{d_A} \left[\openone_{A}
+\sum_{i=1}^{d_A-1}\tau_i^A\gamma_i^A \right]\;.
\end{equation}
The generators form a closed set with respect to commutation: $[\gamma_i^A,\gamma_j^A]=2i\sum_k f_{ijk}\gamma_k^A$, where $f_{ijk}$ is a rank-3 antisymmetric tensor called the structure constant of the group $SU(d_A)$ (see, for example, [G. Mahler and V. A. Weberrus, {\it Dynamics of Open Nanostructures} (Springer Verlag, Berlin, Germany, 1998]). The calculation of commutator (\ref{oz_nec}) is straightforward in this representation, 
\begin{equation}
[\rho,\rho_A\otimes\openone_B]=2i\sum_{h,l,m=1}^{d_A-1}\sum_{k=1}^{d_B-1}
\beta_{hk}\tau_l^A f_{hlm}\gamma_m^A\otimes\gamma_k^B\;.
\end{equation}
Since matrices $\gamma_m^A\;,\gamma_k^B$ are orthogonal, imposing $[\rho,\rho_A\otimes\openone_B]=0$ accounts for constraining  parameters $\beta_{hk}$ and $\tau_l^A$ in the following way,
\begin{equation}
\sum_{h,l=1}^{d_A-1}
\beta_{hk}\tau_l^A f_{hlm}=0
\end{equation}
for all $k$ and $m$. These equations can be inverted. In particular, even the inversion of only one of them is sufficient for our purposes. Doing this, one obtains an explicit  differentiable parametrization of the set $\cz$ with strictly fewer real independent parameters than $d_0$, \ie, the ones required to parametrize $\bqh$. Thus, $\cz$ has Lebesgue measure zero in $\bqh$. $\square$
\section{The set $\cz$ is nowhere dense}
\label{closed_set}

Let us first show that $\cz$ is closed. Since the partial trace is a contractive map -- meaning that the (trace) distance between any two operators is larger than, or equal to, that between the operators resulting from the application of the map --, the map $f:\bqh\longrightarrow f(\bqh)$ is  continuous.  The operator zero (the operator whose matrix representation is composed only of zero elements) in turn forms a closed subset of the set image of $f$, $f(\bqh)$. By the topological  definition of a continuous map, the preimage of a closed set is also closed. Thus, $\cz$ is closed, being the preimage of the closed set  ``operator zero".

To complete the proof, recall that the closure of a closed set is -- by definition -- the set itself. Then a  closed set of  measure zero is nowhere dense because its being measure zero implies that it has no interior point. We show the latter with our example of interest $\cz$: Suppose that there exists an interior point in the closed, zero-measure set  $\cz$. By the definition of interior point, this would mean that there exists a state $\rho\in\cz$ surrounded by an open ball of positive radius entirely contained in $\cz$ (the metric used to define the ball is not relevant, since we are considering finite dimensions). Nevertheless, since open balls have positive Lebesgue measure in ${\mathbb R}^{n}$ for any $n$ this would contradict the fact that $\cz$ has measure zero. Then, there exists no interior point of $\cz$, implying that the set is nowhere dense. $\square$

\section{No finite-time according}
\label{Noaccording}

For any initial state
$\rho$ such that $[\rho,\rho_A\otimes \openone_B]\neq0$, we prove here that there exists no finite time $t_s$ after which the evolved state $\Lambda_t(\rho)$ belongs to   $\cz$ neither for all $t\leq t_s$ nor for $t\in (t_s,t_s+\Delta t]$, with any $\Delta t>0$. Following the notation from the text above, 
we write the condition
 $[\rho_t,{\rho_A}_t\otimes\openone_B]=0\ \forall\ t>t_s$ explicitly as a system of equations 
 to see for their linear independence:
\begin{equation}
\label{commuty}
\sum_{i,j=1}^{d^2}\rho_i\rho_j(\lambda_i\lambda_j)^n\left[\oket{\mu_i},\oket{{\mu_A}_j}\otimes\openone_B\right]=0\ \forall\ n>n_s\in\mathbb{N},
\end{equation}
where $\oket{{\mu_A}_j}=\Tr_B[\oket{\mu_j}]$ and the expansion $\rho=\sum\rho_i \oket{\mu_i}$ has been used. Let us relabel the
pair of indexes $(i,j)$ using a single index $k=1,...,d^2\times d^2$ and define
$R_k=\rho_i\rho_j$, $L_k=\lambda_i\lambda_j$, and
$D_k=\left[\oket{\mu_i},\oket{{\mu_A}_j}\otimes\openone_B\right]$.
Then  Eqs. \eqref{commuty} above can be recast in the form of  linear equations in
$R_k$'s:
\begin{eqnarray}
\sum_k R_k L_k^{n_s+1} D_k &=& 0, \nonumber \\
\vdots \nonumber \\
\sum_k R_k L_k^{n_s+m} D_k &=& 0, 
\label{planeequations}
\end{eqnarray}
for any $m\in\mathbb{N}$, which have to be satisfied conditioned on the initial condition $[\rho,\rho_A\otimes\openone_B]\neq 0$,
\begin{equation}
\label{initialplaneeq}
\sum_k R_k D_k \neq 0.
\end{equation}
We can already intuit  that operator equations \eqref{planeequations} compose a set of $m$ linearly independent equations from the fact 
that coefficients $L_k$ (with $0<|L_k|\leq1$) appear all in a geometric progression. 
We demonstrate this formally by writing  Eqs. \eqref{planeequations} and \eqref{initialplaneeq} in a matrix representation,
 and thus recasting them as a set of linearly independent equations for complex numbers. 

In \refeq{planeequations} and \eqref{initialplaneeq}  we
keep only the $\bar{d}$ terms such that  $D_k$ and $R_k$ are both different
from zero. Thus, 
\begin{eqnarray}\label{app1a}
\sum_{k=1}^{\bar{d}} R_k L_k^{n} D_k &=& 0,\nonumber \\
 \label{app2a}
 \sum_{k=1}^{\bar{d}} R_k D_k &\neq& 0.
\end{eqnarray}
for $n_s<n\leq n_s+m$. Let us now express the operators $D_k$ in an arbitrary matrix
representation and focus on their matrix elements $[D_k]_{p,q}$. The
initial condition \refeq{app2a} implies that there exists at least a
couple $(p_{\bar{0}},q_{\bar{0}})$ such that $\sum_{k=1}^{d'} R_k
[D_k]_{p_{\bar{0}},q_{\bar{0}}} \neq 0$, for some $d'\le\bar{d}$.
Focusing on such a couple $(p_{\bar{0}},q_{\bar{0}})$, and denoting
$d_k\equiv[D_k]_{p_{\bar{0}},q_{\bar{0}}}\neq0$, we have that
Eqs.~(\ref{app1a})) above reduce to ordinary equations
with complex coefficients $d_k$ and  $L_k$:
\begin{eqnarray}
\sum_{k=1}^{d'} R_k L_k^n d_k &=& 0 \;, \\
\sum_{k=1}^{d'} R_k d_k &\neq& 0 \;.
\end{eqnarray}
We now change variables to the non-null coefficients $r_k\doteq R_kd_k$:
\begin{eqnarray}\label{app1}
\sum_{k=1}^{d'} r_k L_k^n &=& 0, \\ \label{app2}
\sum_{k=1}^{d'} r_k &\neq& 0,
\end{eqnarray}
for all $n_s<n\leq n_s+m$. If the coefficients $L_k$ are degenerate, one can define
another set of variables by grouping together all the $r_k$'s that correspond to
the same degenerate $L_k$. Namely, we introduce $s_h=\sum r_k$, where the
sum extends to the $r_k$'s corresponding to the same $L_h$. Denoting by
$\tilde{d}$ the number of different $L_h$'s we have that
Eqs.~(\ref{app1}) and (\ref{app2}) are equivalent to:
\begin{eqnarray}
\label{finaleqs}
\sum_{h=1}^{\tilde{d}} s_h L_h^{n_s+1} &=& 0
 \nonumber \\
\vdots \nonumber \\
\sum_{h=1}^{\tilde{d}} s_h L_h^{n_s+m} &=& 0, \\ \label{app4}
\sum_{h=1}^{\tilde{d}} s_h &\neq& 0,
\end{eqnarray}
with $L_{h}\neq L_{h'}$ if $h\neq h'$ and $n_s<n\leq n_s+m$. 
Equations \eqref{finaleqs} are linear in $s_h$ with complex, non-null,
non-degenerate coefficients in geometric progression. From the properties of eigenvalues $\lambda_i$ mentioned in the text, 
we see that coefficients $L_h$ necessarily satisfy $|L_h|\leq1\ \forall\ h$, with $|L_h|=1$ for  some $h$ and  $|L_h|<1$ for 
all other $h$'s.  Thus, Eqs. \eqref{finaleqs} yield an 
homogenous system of $m$ independent linear equations for $\tilde{d}$ unknowns  $s_h$. For $m<\tilde{d}$ 
there are $m$ nontrivial solutions that are also compatible with \eqref{app4}. For $m\geq\tilde{d}$ though, Eqs. \eqref{finaleqs} 
become a uniquely-determined homogenous system, whose unique solution is the trivial one  $s_h=0$ for all $h=1,...,\tilde{d}$.  
This solution, however, is not acceptable, since it contradicts the initial
condition \refeq{app4}. $\square$

As said,  trajectories that cross  $\cz$ at most  $\tilde{d}-1$ times might in principle give acceptable solutions to Eqs. \eqref{finaleqs}.
If there are $\tilde{d}_{\lambda}$ different $\lambda_i$-eigenvalues, it is straightforward to count that there are
$\tilde{d}=\tilde{d}_{\lambda}(\tilde{d}_{\lambda}-1)/2$ different $L_h$'s. As an example, we can now easily calculate an upper bound to the number of  times
 $\cz$ can be crossed by usual maps, such as a local depolarizing or dephasing channels (three different eigenvalues, two times), or
the global depolarizing channel (two different eigenvalues, never). 

 \par On the other hand, also from Eqs. \eqref{finaleqs} one can see that $\rho_t\in\cz$, for $t\rightarrow\infty$ if, and only if, the steady state of the map is itself a state inside $\cz$. This is clear when one considers the limit $n_s\rightarrow\infty$ in  \eqref{finaleqs}, where all powers of $L_h$, from $L_h^{n_s+1}$ to $L_h^{n_s+m}$, are exactly equal to zero for $|L_h|<1$, and equal to 1 for the single $L_h$ equal to one. Eqs. \eqref{finaleqs} simply converge to the single condition $s_{H}=0$, where $H$ is the one $h$ for which  $L_{H}=1$. This condition is in turn not in conflict with \eqref{app4} and therefore provides an acceptable solution. The coefficient  $s_{H}$ is associated to the projection of the initial state onto the map's steady state. So it simply gives the trivial fact that the final state will end up in $\cz$ if and only if the steady state of the map is itself in $\cz$.


\begin{thebibliography}{99}
\bibitem{note_other}
\par  Notice that these properties automatically extend also to ``two-side measures" of quantumness \cite{Luo&Wu} based on the disturbance caused by local measurements on both subsystems \cite{Terhal&DiVincenzo,PHH}.
This is because discord yields a general lower bound to such measures \cite{Luo&Wu}, for the disturbance when measuring both particles is never less than that caused when only one particle is subject to measurements.
\end{thebibliography}
\end{document}